\documentclass[prd,showpacs]{revtex4}
\usepackage{amsmath}
\usepackage{graphicx}
\usepackage{epsfig}
\usepackage{dcolumn}
\usepackage{bm}
\usepackage{dcolumn}
\usepackage{slashed}
\usepackage{amsfonts}
\usepackage{amssymb}
\providecommand{\U}[1]{\protect\rule{.1in}{.1in}}
\providecommand{\U}[1]{\protect\rule{.1in}{.1in}}
\providecommand{\U}[1]{\protect\rule{.1in}{.1in}}
\providecommand{\U}[1]{\protect\rule{.1in}{.1in}}
\providecommand{\U}[1]{\protect\rule{.1in}{.1in}}

\begin{document}

\title{Can the particle mass spectrum be explained within the Standard Model?}
\author{A. Cabo Montes de Oca$^{*\, ,\,**}$, N. G. Cabo-Bizet$^{***\, ,\,****}$ and A.
Cabo-Bizet$^{***}$}

\affiliation{$^{*}$\textit {\ Perimeter Institute for Theoretical
Physics, 31 Caroline St. N., Waterloo, Ontario, Canada }}
\affiliation{$^{**}$\textit{\ Grupo de F\'{\i }sica Te\'orica,
Instituto de Cibern\'etica, Matem\'atica y F\'{\i}sica, Calle E, No.
309, Vedado, La Habana, Cuba}}
\affiliation{$^{***}$\textit{Departamento de F\'{\i}%
sica, Centro de Aplicaciones Tecnol\'{o}gicas y Desarrollo Nuclear
(CEADEN), Calle 30, esq. a 5ta Ave, La Habana, Cuba. }}
\affiliation{$^{****}$ \textit {Institute of Physics, Bonn
University, Nussalee 12, 53115, Bonn, Germany}}

\begin{abstract}
\noindent A modified version of PQCD considered in previous works is
investigated here in the case of retaining only the quark
condensate. The Green functions generating functional is expressed
in a form in which Dirac's delta functions are now absent from the
free propagators. The new expansion implements the dimensional
transmutation effect through a single interaction vertex in addition
to the standard ones in mass less QCD. The new vertex suggest a way
for constructing an alternative to the SM in which the mass and CKM
matrices could be generated by the instability of masslesss QCD
under the production of the top quark and other fermions
condensates, in a kind of generalized Nambu Jona Lasinio mechanism.
The results of a two loop evaluation of the vacuum energy indicate
that the quark condensate is dynamically generated. However, the
energy as a function of the condensate parameter is again unbounded
from below in this approximation. Assuming the existence of a
minimum of the vacuum energy at the experimental value of the top
quark mass $m_{q}=173$ $GeV$, we evaluate the two particle
propagator in the quark anti-quark channel in zero order in the
coupling and a ladder approximation in the condensate vertex.
Adopting the notion from the former $top$ quark models, in which the
Higgs field corresponds to the quark condensate, the results
suggests that the Higgs particle could be represented by a meson
which might appear at energies around two times the top quark mass.
\end{abstract}


 \maketitle

\section{Introduction}

The origin of the singular structure of the particle mass spectrum is one of
the central questions in \ Particle Physics. \ \ Although the Standard Model
(SM)\ furnishes a remarkable description of the physical experience, the issue
about better understanding the \ mass hierarchy has been always present in the
research activity \cite{nambu,fritzsch,coleman,
bardeen,miransky,minkowski,clague}. In particular, this circumstance is
reflected in the unsatisfying large \ number of parameters which should be
fixed in the SM to describe the observed masses. Therefore, the search of
\ new approaches to consider this problem is an important theme of study
nowadays. \ \ \ \

A development of \ an alternative perturbation expansion for QCD, \
including the presence of quark and gluon condensates in the free
vacuum state generating the Wick expansion, has been considered in
previous works
\cite{mpla,prd,epjc,epjc1,jhep,epjc2,hoyer,hoyer1,hoyer2,epjc19}. \
A basic issue motivating the study is the question about what could
be the final strength of a \ dynamically generated \ quark
condensate in mass less QCD, in which the free vacuum is strongly
degenerated \ and the underlying forces are the strongest ones in
Nature \cite{coleman}. \ This point motivated the search for
modifications of the Wick expansion in this theory starting from
free vacuum states including zero momentum gluon and quark
condensates, in order to afterwards adiabatically connect the
interaction \cite{prd,jhep,epjc}. \ A similar modification of the
free vacuum leading to the perturbative expansion, but filling real
particles states up to a Fermi level was before considered by P.
Hoyer \cite{hoyer}. \ \ The modified expansion following from Ref.
\cite{mpla,prd,epjc,epjc1,jhep,epjc2,epjc19} has also close
connections with other independent approaches in the literature.
Those consider modified free particle vacua and propagators in
generating the perturbative expansion in order to take into account
condensate effects \cite{celenza, roberts, pavel}. Finally, we
expect that in the future the analysis can show links with well
established non perturbative theories considering condensation
effects such as the Sum Rules approach and the Fukuda gluon
condensation studies \cite{shifman,fukuda}. We would like to remark
that the modified expansion being investigated could perhaps
represent  a theoretical foundation for the "superconductivity
systems " like properties of the particle mass spectrum underlined
in Refs. \cite{nambu,fritzsch, fritzsch1,bardeen}. This possibility
is signaled by the fact that the fermion condensate generation
closely resembles the similar effect in the usual BCS theory. This
fact is a natural outcome since the free vacua employed to generate
the expansion have the same BCS like "squeezed" state
structure\cite{prd}.
 \ In
Ref.\cite{mpla,prd,epjc,epjc1,jhep,epjc2,epjc19} some indications
about the possible dynamic generation of quark and gluon condensates
had been obtained. Nevertheless, this correction to the vacuum
energy turned out to be unbounded from below as a function of the
quark condensate. \ \ In this work \ we \ restrict the discussion to
the simpler case in which there is only one quark condensate present
in the system. It is natural to\ firstly consider this \ situation,
since the aim is to investigate the possibility that \ large dynamic
quark condensates and masses could define a kind of $top$ quark
model \ as an effective action for mass less QCD. \ In this \ case,
the Green' functions generating functional $Z$ of the system
obtained in Ref.\cite{epjc19}, is here transformed to a more helpful
representation. In this form $Z$ is expressed as the same functional
integral associated \ to mass less \ QCD, in which all the effects
of the condensates are now embodied in only one special \ vertex \
having two quark and two gluon legs. This representation allows to
systematize the \ diagrammatic expansion of the problem. In
particular it permits to \ implement the dimensional transmutation
effect. \ \ \ The obtained path integral formula \ can be expected
to be also helpful in developing perturbative schemes in
superconductivity theory. \ This formula is perhaps the central
result of the present work, because it indicates  a  technical path
through which the approach being considered, could help to\ evidence
that \ an strong instability of massless QCD under the generation of
fermion condensates,  can be the explanation  of the whole particle
mass hierarchy (including quark and leptons)  in \ a sort of \
generalized \ Nambu-Jona Lasinio SSB mechanism \cite{nambu,fritzsch,
bardeen,miransky}. Specifically,  the structure of the vertex
indicates ways for its generalization \ which seem able to describe
the quark mass and CKM matrices as coming from the first terms in an
effective action.

The application of the expansion is  \cite{nambu,fritzsch,coleman,
bardeen,miransky,minkowski,clague} considered in this work, by
calculating the leading logarithm of the condensate dependence of a
two loop approximation considered for the effective potential. The
results repeat the indication of the dynamic generation of the quark
condensate obtained in previous works \cite{epjc2,epjc19}, but
again, the potential results to be unbounded from below. \ However,
in the present case, the obtained leading logarithm behavior
reinforces the instability for large values of the quark condensate.
This\ outcome rises the need of performing new evaluations that
could determine a minimum. The attainment of stability in the next
three loop approximation is \ feasible, since \ squared logarithms
of the condensate terms should appear, that upon showing the
appropriate sign can produce a global minimum of the potential. \
The evaluation of the squared logarithm corrections at the\ three
loop level is expected to be considered elsewhere.

The work also present an evaluation of the two particle propagator in a
$t\overline{t}$ channel in zero order in the coupling and a ladder
approximation in the condensate vertex. The singularities of the result are
then analyzed by assuming the existence of a minimum of the vacuum energy at
the experimental value of the top quark mass $m_{q}=173$ GeV. \ In this case,
after also adopting the notion from the former $top$ quark models, in which
the Higgs field corresponds to the quark condensate, the results suggest that
the Higgs particle should be considered as a $t\overline{t}$ meson which could
appear at energies\ around to two times the $top$ quark mass. \ The \ mass of
this meson bound state is expected to be estimated after adding the gluon
exchange contribution to the kernel of the Bethe-Salpeter equation, to the
here evaluated only condensate dependent kernel.

\ The work proceeds as follows. \ In Section II the function
integral formula for the states showing a quark condensate \ is
derived. \ \ Section III is devoted to evaluate two correction for
the vacuum energy as \ function of the condensate parameter. \ The
Section IV \ then consider the evaluation of the two particle Green
function in ladder approximation in terms of the new condensate
vertex and the zero order in the coupling. Finally the results are
reviewed in the Summary.

\section{A functional integral for quark condensate states}

In this section we will \ present a simpler representation of the
generating functional of the modified massless QCD \ in which a
fermion condensate is introduced to define the initial vacuum state
employed to generate the Wick expansion \cite{epjc19}. \ The
unrenormalized form of the functional will be considered. In this
case all the condensate parameters are absent from the vertices. The
expression\ for the  complete generating functional introduced in
Ref. \cite{epjc19} as
restricted to a vanishing gluon condensate can be written in the form%
\begin{align}
Z[j,\xi,\xi^{\ast},\eta,\overline{\eta}]  &  =\exp[i\int dx\int\mathcal{L}%
_{1}(\frac{\delta}{i\delta j^{a,\mu}},\frac{\delta}{i\delta\xi^{\ast}}%
\frac{\delta}{-i\delta\xi}\frac{\delta}{i\delta\overline{\eta}}\frac{\delta
}{-i\delta\eta})]\times\label{vertex}\\
&  Z^{(0)}[j,\xi,\xi^{\ast},\eta,\overline{\eta}],\nonumber\\
Z^{(0)}[j,\xi,\xi^{\ast},\eta,\overline{\eta}]  &  =Z^{G}[j]Z^{FP}[\xi
,\xi^{\ast}]Z^{F}[\eta,\overline{\eta}],\\
\mathcal{L}_{1}(A_{\mu},\chi,\chi,\psi,\overline{\psi})  &  =-\frac{g}%
{2}f^{abc}(\partial_{\mu}A_{\nu}-\partial_{\nu}A_{\mu})A^{b\mu}A^{c\nu}%
-g^{2}f^{abe}f^{cdec}A_{\mu}^{a}A_{\nu}^{b}A^{c\mu}A^{d\nu},-\nonumber\\
&
-gf^{abc}\partial^{\mu}(\chi^{a\ast})\chi^{b}A_{\mu}^{c}+g\overline{\psi
}T^{a}\gamma^{\mu}\psi A_{\mu}^{a},
\end{align}
where the free generating functionals associated \ the gluon, ghosts and quark
fields take the expressions%

\begin{align*}
Z^{G}[j]  &  =\exp[\frac{i}{2}\int dx\text{ }dy\text{ }j^{a\mu}(x)D_{\mu\nu
}^{ab}(x-y)\text{ }j^{b\nu}(x)],\\
Z^{FP}[\xi,\xi^{\ast}]  &  =\exp[i\int dx\text{ }dy\text{ }\xi^{a\ast
}(x)D(x-y)\text{ }\xi^{a}(x)],\\
Z^{F}[\eta,\overline{\eta}]  &  =\exp[i\int dx\text{ }dy\text{ }\overline
{\eta}(x)S^{C}(x-y)\eta(y)].
\end{align*}

The exponential operator $\exp[i\int
dx\int\mathcal{L}_{1}(\frac{\delta }{i\delta
j^{a,\mu}},\frac{\delta}{i\delta\xi^{\ast}}\frac{\delta}{-i\delta
\xi}\frac{\delta}{i\delta\overline{\eta}}\frac{\delta}{-i\delta\eta})]$
will be denominated in what follows as the $vertex$ $part$. By the
assumption of vanishing \ gluon condensate, the gluon and ghost
propagators are the usual
Feynman ones and the quark propagator includes the condensate dependent part as%

\[
S^{C}(x-y)=\int\frac{dk}{(2\pi)^{D}}(\frac{1}{-p_{\mu}\gamma^{\mu}}-i\text{
}C\text{ }\delta^{D}(p))\exp(-i\text{ }p.x).
\]

Let us repeat below, for this simpler case of \ only having the
quark condensate, the procedure employed in \cite{epjc19} for
linearizing in the sources the exponential arguments in the
generating functional.  The free quark generating functional can be
rewritten \ in the form
\[
Z^{F}[\eta,\overline{\eta}]=\exp[i\int dxdy\text{ }\overline{\eta
}(x)S(x-y)\eta(y)]\times\exp[\int dx\text{ }\overline{\eta}(x)\frac{C}%
{(2\pi)^{D}}\int dy\text{ }\eta(y)],
\]
and the quadratic in the sources argument of the exponential can be
represented as a linear one after expressing the exponential as the
result of the gaussian integral
\begin{align*}
\exp[\int dx\text{ }\overline{\eta}(x)\frac{C}{(2\pi)^{D}}\int dy\text{ }%
\eta(y)] &  =\int\mathcal{D}\overline{\chi}\mathcal{D}\chi\exp[-\overline
{\chi}\chi+\\
&  +i{\Large (}\int dx\text{ }\overline{\eta}(x)\frac{1}{i}(\frac{C}%
{(2\pi)^{D}})^{\frac{1}{2}}\chi+\overline{\chi}\frac{1}{i}(\frac{C}{(2\pi
)^{D}})^{\frac{1}{2}}\int dy\text{ }\eta(y){\Large )].}%
\end{align*}

As a consequence of the \ implemented linearity in the sources, the
condensate parameter $C$ dependent terms can be shifted to the left
of the $vertex$ $part$ functional operator in equation
(\ref{vertex}) by employing the
following general relation%

\begin{align*}
&  \mathcal{F[}\frac{\delta}{i\delta\overline{\eta}},\frac{\delta}%
{-i\delta\eta}]\exp[i{\Large (}\int dx\text{ }\overline{\eta}(x)\frac{1}%
{i}(\frac{C}{(2\pi)^{D}})^{\frac{1}{2}}\chi+\overline{\chi}\frac{1}{i}%
(\frac{C}{(2\pi)^{D}})^{\frac{1}{2}}\int dy\text{ }\eta(y){\Large )]}\\
&  =\exp[i{\Large (}\int dx\text{ }\overline{\eta}(x)\frac{1}{i}(\frac
{C}{(2\pi)^{D}})^{\frac{1}{2}}\chi+\overline{\chi}\frac{1}{i}(\frac{C}%
{(2\pi)^{D}})^{\frac{1}{2}}\int dy\text{ }\eta(y){\Large )]\times}\\
&  \mathcal{F[}\frac{\delta}{i\delta\overline{\eta}}+\frac{1}{i}(\frac
{C}{(2\pi)^{D}})^{\frac{1}{2}}\chi,\frac{\delta}{-i\delta\eta}+\overline{\chi
}\frac{1}{i}(\frac{C}{(2\pi)^{D}})^{\frac{1}{2}}].
\end{align*}

Then, after again representing as functional integrals, the free
generating functionals associated to the gluons, ghosts and quarks,
by also acting on them \ with the new terms  appeared in the
$vertex$ $part$ after the above described commutations,  a modified
free theory generating functional can be written in the following
way \cite{epjc19}
\begin{align}
Z^{(0,C)}[j,\eta,\overline{\eta},\xi,\overline{\xi}|C_{q}] &  =\frac
{1}{\mathcal{N}}\int\int\int d\overline{\chi}d\chi\mathcal{D}[A,\overline
{\Psi},\Psi,\overline{c},c]\exp[i\text{ }S^{(0)}[A,\overline{\Psi}%
,\Psi,\overline{c},c,\overline{\chi},\chi]]\nonumber\\
&  =\frac{1}{\mathcal{N}}\int\int d\overline{\chi}d\chi\exp[-\overline{\chi
}_{u}^{i}\chi_{u}^{i}]\int\mathcal{D}[A,\overline{\Psi},\Psi,\overline
{c},c]\times\nonumber\\
&  \times\exp[-i\int\frac{dk}{(2\pi)^{D}}[\frac{1}{2}A_{\mu}^{a}%
(-k)(k^{2}g^{\mu\nu}-(1-\frac{1}{\alpha})k_{\mu}k_{\nu})A_{\nu}^{a}%
(k)+\nonumber\\
&  i\int\frac{dk}{(2\pi)^{D}}\overline{c}^{a}(-k)k^{2}c^{a}(k)+\nonumber\\
&  +i\int\frac{dk}{(2\pi)^{D}}\overline{\Psi}^{i}(-k)\text{ }\gamma_{\mu
}k^{\mu}\,\Psi^{i}(k)\text{ }+\nonumber\\
&  +i\int\frac{dk}{(2\pi)^{D}i}\overline{\Psi}^{i,u}(-k)\text{ }g(\frac{C_{q}%
}{(2\pi)^{D}})^{\frac{1}{2}}\gamma_{\mu}^{u\text{ }v}T_{a}^{ij}\,\chi
^{j,v}\text{ }A^{\mu,a}(k)+\nonumber\\
&  +i\int\frac{dk}{(2\pi)^{D}i}A^{\mu,a}(-k)\overline{\chi}^{i,u}\text{
}g(\frac{C_{q}}{(2\pi)^{D}})^{\frac{1}{2}}\text{ }\gamma_{\mu}^{u\text{ }%
v}T_{a}^{ij}\,\Psi^{j,v}(k)+\label{Z0}\\
&  +i\int\frac{dk}{(2\pi)^{D}}(j_{\mu}(-k)A^{\mu}(k)+\overline{\eta}%
(-k)\Psi(k)+\overline{\Psi}(k)\eta(-k)+\nonumber\\
&  \overline{\xi}(-k)c(k)+\overline{c}(-k)\xi(k))].\nonumber
\end{align}

It corresponds to the general relation obtained  \cite{epjc19} \
after taking a vanishing gluon condensate parameter. The complete
generating functional is obtained by acting on it with the usual
$vertex$ $part$ functional operator.  It should be noted that the
above expression has been written in its Minkowski space form, \
since in this work we will adopt the same conventions and notations
as in Ref. \cite{muta}. In the formula $A$, $\Psi$ \ and $c$ \ are
the gluon, quark and ghost fields and $\overline{\chi},\chi$ are the
space independent auxiliary parameters which were introduced in Ref.
\cite{epjc19} in order to represent the quadratic forms in the
sources as linear ones. The mentioned in the Introduction
simplification of the perturbative expansion, \ comes from noticing
that in \ expression (\ref{Z0}) \ the integral over the auxiliary
fields is a Gaussian one. \ Therefore, it can be explicitly
integrated by finding the values of the auxiliary fields which solve
the Lagrange equations following from the action laying in the
argument of the exponential integrand. \ These equations of
motion take the simple forms%
\begin{align}
\frac{\delta S^{(0)}[A,\overline{\Psi},\Psi,\overline{c},c,\overline{\chi
},\chi]}{\delta\overline{\chi}_{u}^{i}} &  =-\chi_{u}^{i}+i\int\frac{dk}%
{(2\pi)^{D}i}A^{\mu,a}(-k)g(\frac{C_{q}}{(2\pi)^{D}})^{\frac{1}{2}}\text{
}\gamma_{\mu}^{u\text{ }v}T_{a}^{ij}\,\Psi^{j,v}(k)=0,\label{L1}\\
\frac{\delta S^{(0)}[A,\overline{\Psi},\Psi,\overline{c},c,\overline{\chi
},\chi]}{\delta\chi_{u}^{i}} &  =\overline{\chi}_{u}^{i}-i\int\frac{dk}%
{(2\pi)^{D}i}\overline{\Psi}^{j,v}(-k)\text{ }g(\frac{C_{q}}{(2\pi)^{D}%
})^{\frac{1}{2}}\gamma_{\mu}^{vu\text{ }}T_{a}^{ji}\,\text{ }A^{\mu
,a}(k)=0.\label{L2}%
\end{align}

Henceforth, after substituting the expressions for the \ auxiliary fields \ in
\ \ equation (\ref{Z0}),\ the free generating functional can be written as
follows%
\begin{align}
Z^{(0)}  &  =\frac{1}{\mathcal{N}}\int\mathcal{D}[A,\overline{\Psi}%
,\Psi,\overline{c},c]\exp[iS^{(0)}[A,\overline{\Psi},\Psi,\overline
{c},c]+i\text{ }S^{(C_{q})}[A,\overline{\Psi},\Psi]],\label{Z01}\\
\text{ }S^{(0)}[A,\overline{\Psi},\Psi,\overline{c},c]  &  =\left.
S^{(0)}[A,\overline{\Psi},\Psi,\overline{c},c,\overline{\chi},\chi]\right\vert
_{\overline{\chi},\chi=0},
\end{align}
where the two terms linear in the auxiliary fields have been substituted by
the action $S^{(C_{q})}$ , having the expression%
\begin{align}
S^{C_{q}}[A,\overline{\Psi},\Psi]  &  =\frac{g^{2}C_{q}}{i(2\pi)^{D}}\int
\frac{dk}{(2\pi)^{D}}\overline{\Psi}^{i,u}(-k)\gamma_{\mu}^{uv}T_{a}%
^{ij}A^{\mu,a}(k)\times\nonumber\\
&  \int\frac{dk^{\prime}}{(2\pi)^{D}}A^{\mu^{\prime},a^{\prime}}(-k^{\prime
})\gamma_{\mu^{\prime}}^{vu^{\prime}}T_{a}^{ji^{\prime}}\Psi^{i^{\prime
}u^{\prime}}(k^{\prime})\nonumber\\
&  =\frac{g^{2}C_{q}}{i(2\pi)^{D}}\int\int dxdx^{\prime}\overline{\Psi}%
^{i,u}(x)\gamma_{\mu}^{uv}T_{a}^{ij}A^{\mu,a}(x)A^{\mu^{\prime},a^{\prime}%
}(x^{\prime})\gamma_{\mu^{\prime}}^{vu^{\prime}}T_{a^{\prime}}^{ji^{\prime}%
}\Psi^{i^{\prime}u^{\prime}}(x^{\prime})\nonumber\\
&  =\frac{g^{2}C_{q}}{i(2\pi)^{D}}\int\int dxdx^{\prime}\overline{\Psi
}(x)\slashed{A}(x)\slashed{A}(x^{\prime})\Psi^{^{\prime}}(x^{\prime}).
\label{Vc}%
\end{align}

\begin{figure}[h]
%\centering
%
\includegraphics[width=6cm]{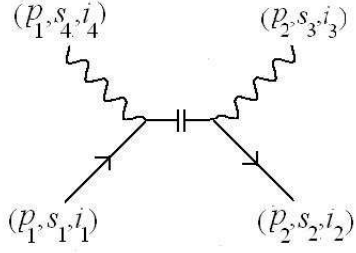}\caption{{The diagram shows the structure
of the new vertex which must be added to the Feynman diagram rules of mass
less QCD to describe the modified Wick expansion of the theory in presence of
a quark condensate in the free vacuum.}}%
\label{vertex}%
\end{figure}

In the last line of the equation, $\overline{\Psi}^{i,u}(x)$ and
$\Psi^{i^{\prime}u^{\prime}}$ are understood as vectors with composite indices
formed by the color and spinor ones and $\slashed{A}$ should correspondingly
be interpreted as a matrix with such kind of indices also. This term defines a
new interaction vertex \ which convey all the information about the fermion
condensate. \ \ \ By adding now the action terms associated to the other usual
interaction vertices of mass less QCD, the full expression for the generating
functional of the modified PQCD can be \ written in the form%

\begin{equation}
Z[j,\eta,\overline{\eta},\xi,\overline{\xi}|C_{q}]=\frac{1}{\mathcal{N}}%
\int\mathcal{D}[A,\overline{\Psi},\Psi,\overline{c},c]\exp[i\text{
}S[A,\overline{\Psi},\Psi,\overline{c},c]+i\text{ }S^{C_{q}}[A,\overline{\Psi
},\Psi]], \label{Z}%
\end{equation}
in which now $S$ is the full action defining mass less QCD \cite{muta}:%
\begin{align}
S  &  =\int dx(\mathcal{L}_{0}\mathcal{+L}_{1}\mathcal{)},\nonumber\\
\mathcal{L}_{0}  &  =\mathcal{L}^{g}+\mathcal{L}^{gh}+\mathcal{L}%
^{q},\nonumber\\
\mathcal{L}^{g}  &  =-\frac{1}{4}(\partial_{\mu}A_{\nu}^{a}-\partial_{\nu
}A_{\mu}^{a})(\partial^{\mu}A^{a,\nu}-\partial^{\nu}A^{a,\mu})-\frac
{1}{2\alpha}(\partial_{\mu}A^{\mu,a})(\partial^{\nu}A_{\nu}^{a}),\nonumber\\
\mathcal{L}^{gh}  &  =(\partial^{\mu}\chi^{\ast a})\partial_{\mu}\chi
^{a},\label{S}\\
\mathcal{L}^{q}  &  =\overline{\Psi}(i\gamma^{\mu}\partial_{\mu}%
)\Psi,\nonumber\\
\mathcal{L}_{1}  &  =-\frac{g}{2}f^{abc}(\partial_{\mu}A_{\nu}^{a}%
-\partial_{\nu}A_{\mu}^{a})A^{b,\mu}A^{c,\nu}-g^{2}f^{abe}f^{cde}A_{\mu}%
^{a}A_{\nu}^{b}A^{c,\mu}A^{d,\nu}-\nonumber\\
&  -gf^{abc}(\partial^{\mu}\chi^{\ast a})\chi^{b}A_{\mu}^{c}+g\overline{\Psi
}T^{a}\gamma^{\mu}\Psi A_{\mu}^{a}.\nonumber
\end{align}
\

At this point we would like to comment about the possibility for a natural way
of generalizing the form of the new vertex to introduce the interactions
between quarks and leptons of different flavors. \ The proposal has the form%
\begin{equation}
S^{C}[A,\overline{\Psi},\Psi]=\sum\limits_{f_{1},f_{2}}^{\sigma_{1},\sigma
_{2}}\frac{C_{f_{1}f_{2}}^{\sigma_{1}\sigma_{2}}}{i(2\pi)^{D}}\int\int
dxdx^{\prime}\overline{\Psi}_{f_{1},\sigma_{1}}(x)\slashed{A}(x)\slashed
{A}(x^{\prime})\Psi_{f_{2},\sigma_{2}}(x^{\prime}),\label{Gen}%
\end{equation}
where $f_{1},f_{2}$ are quark   flavor indices and
$\sigma_{1},\sigma_{2}$ are weak interactions $SU(2)$ ones. Note
that this structure could be even more generalized to include the
six lepton flavors. The \ discussion in Ref. \cite{jhep}, by
example, \ directly suggests that the Feynman expansion \ generated
have the chance of allowing to \ phenomenologically reproduce the
quark mass and CKM matrices of the SM. This possibility in
conjunction with the definition of a ground state in which the
condensate stabilizes, \ can give also add support to the \
alternative expansion being investigated. \ These issues are
expected to be considered elsewhere. \ \

\section{Two loop leading logarithm correction to the Effective Potential}

Let us in this section consider the vacuum energy (the negative of the
Effective Potential) as a function of the \ condensate parameter. The general
motivation is to investigate \ the possibility for obtaining a \ minimal
energy vacuum state around which the systems stabilizes. This stabilization,
then, can open the way for the interest to investigate the physical
predictions of the perturbation expansion around this stable state. In
particular the most ambitious expectation is the possibility for generating
the \ physics of the SM from a generalized version of the scheme after
including the rest of the fields needed to such an objective.
\begin{figure}[h]
%\centering
%
\includegraphics[width=8cm]{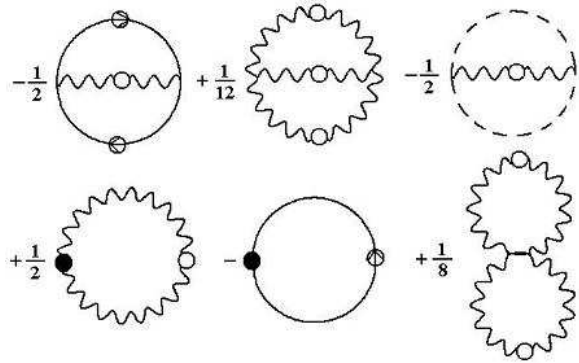}\caption{ The two loop correction to the
Effective Action considered in this work. It corresponds to the sum of the
same two loop diagrams of mass less QCD, but in which the free mass less quark
and gluon propagators were substituted by \textit{dressed} counterparts.}%
\label{twoloop}%
\end{figure}

\begin{figure}[h]
%\centering
%
\includegraphics[width=8cm]{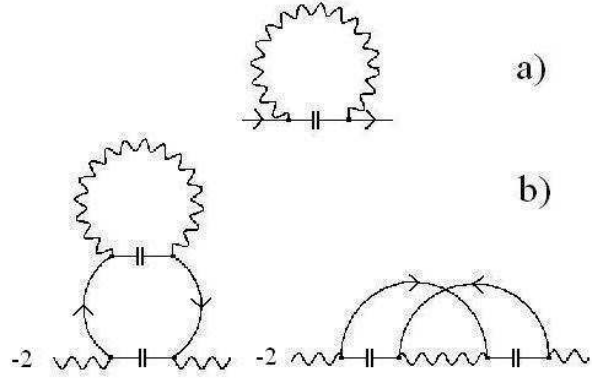}\caption{ Figure a) shows the one loop
quark self-energy diagram employed for constructing the \textit{dressed} quark
propagator in the ladder approximation. Figure b) illustrates the two loop
self-energy correction employed to define the gluon \textit{dressed}
propagator also in the ladder approximation. A two loop correction was
required because the one loop one vanishes. For this exploration, these
corrections were chosen to only depend on the new condensate vertex. }%
\label{selfenergy}%
\end{figure}

The specific approximation to be taken for the Effective Potential will be the
following one. We will consider the same summation of two loop graphs of mass
less QCD, but substituting the \ mass less free quark and gluon propagators by
$dressed$ expressions. These ones will be propagators associated to
self-energies taken in their lowest non vanishing order in terms of the
condensate dependent vertex defined in (\ref{Vc}) and Fig. \ref{vertex}. \ The
two loop contributions are shown in Fig. \ref{twoloop}. In this picture the
above mentioned quark and gluon propagators are indicated by the usual wavy
and straight lines respectively, but adding open circles their mid points. The
large black dots denote the counterterm vertices of QCD appearing in the
considered two loop approximation. Further below the appearing \ gluon and
quark lines without circles will mean the usual free propagators \ of mass
less QCD in the conventions of Ref. \cite{muta} \ The $dressed$ quark
propagator \ will \ be the one generated by the infinite ladder of insertions
of the one loop self energy illustrated in Fig \ref{selfenergy}a) . \ For
\ the gluons, the fact that the one loop self-energy (polarization operator)
contribution vanishes, requires to consider the two loop gluon self energy
terms which Feynman diagrams are shown in Fig \ref{selfenergy}b). \ At this
point it could be useful to recall that the evaluations considered in the
paper were done by following the notations and conventions of \ Ref.
\ \cite{muta}. As above noted, the \ Feynman graph and its analytic expression
of the new vertex to be added to the standard ones in Ref. \cite{muta} are
defined in Fig. 1 and Eq. (\ref{Vc}). 

The \ gluon\ and quark self-energy evaluations in Fig. \ref{selfenergy} need
only for the calculation of some spinor and color traces, since the loop
integrals \ are canceled in both cases by the Dirac's Delta functions
appearing in the condensate dependent vertex. The results for them has the
expressions%
\begin{align}
\Sigma_{i,j}^{u,v}(p)  &  =-S\frac{\delta^{ij}\delta^{uv}}{p^{2}%
},\label{sigma}\\
S  &  =-\frac{g^{2}C_{q}}{(2\pi)^{D}}\frac{D(N^{2}-2)}{2N},\label{defS}\\
\Pi_{\mu\nu}^{ab}(p)  &  =-\frac{\delta^{ab}}{(p^{2})^{2}}{\large (}a\text{
}(g_{\mu\nu}-\frac{p_{\mu}p_{\nu}}{p^{2}})+(a+b)\frac{p_{\mu}p_{\nu}}{p^{2}%
}{\large )},\label{pimunu}\\
a  &  =S^{2}\frac{8N(DN^{2}+4-2D)}{D^{2}(N^{2}-1)^{2}},\text{ \ }b=S^{2}%
\frac{16N(D-2)}{D^{2}(N^{2}-1)^{2}}, \label{ayb}%
\end{align}
where $D=4-2\epsilon$ is the space dimension in the dimensional regularization
scheme and $N=3$ defines the $SU(3)$ group of interest here.

When the gauge $\alpha=1$ is chosen, as it will done in this work, the
summations of the \ geometric series representing all the ladder self-energy
insertions leads for the quark propagator the expression
\begin{align}
G_{i_{1}i_{2}}^{u_{1}u_{2}}(p)  &  =\delta^{i_{1}i_{2}}{\Large (}\frac
{1}{-p_{\mu}\gamma^{\mu}+\frac{S}{p^{2}}}{\Large )}^{u_{1}u_{2}}\nonumber\\
&  =-\frac{\delta^{i_{1}i_{2}}}{p^{2}-\frac{S^{2}}{(p^{2})^{2}}}%
{\Large (}p_{\mu}\gamma^{\mu}+\frac{S}{p^{2}}{\Large )}^{u_{1}u_{2}%
},\label{quarprop}\\
S  &  =-m_{q}^{3}, \label{mqdef}%
\end{align}
the quantity $m_{q}$ defines the only pole laying in the positive real axis of
the propagator as a function of $\ p^{2}$. It will named in what follows as
the \ $quark$ $mass$. As it was discussed in Ref. \cite{epjc19} the
\ parameters of \ perturbative expansion may be chosen to be \ $g$ and $m_{q}$ .

In the case of the gluon propagator, after the summation of the geometric
series associated to the sum of arbitrary insertions of the gluon self-energy
illustrated in Fig. \ref{selfenergy}b), the $dressed$ gluon propagator can be
written in the form
\begin{align}
D_{\mu\nu}^{ab}(p)  &  =D_{\mu\nu}^{(0)ab}(p)+D_{\mu\nu}^{(1)ab}(p)+D_{\mu\nu
}^{(2)ab}(p),\\
D_{\mu\nu}^{(0)ab}(p)  &  =\frac{\delta^{ab}g_{\mu\nu}}{p^{2}},\\
D_{\mu\nu}^{(1)ab}(p)  &  =-\frac{a\delta^{ab}g_{\mu\nu}}{p^{2}((p^{2}%
)^{3}+a)},\\
D_{\mu\nu}^{(2)ab}(p)  &  =-\frac{b\delta^{ab}p^{2}p_{\mu}p_{\nu}}%
{((p^{2})^{3}+a)((p^{2})^{3}+(a+b))}, \label{gluonprop}%
\end{align}
in which $D_{\mu\nu}^{(0)ab}$ is the free propagator and $D_{\mu\nu}%
^{(1)ab}(p)$ and $D_{\mu\nu}^{(2)ab}(p)$ the condensate dependent
contributions which vanish if $C_{q}=0.$

Now, it can be observed that\ after removing the dimensional regularization,
any Feynman graphs contributing to the vacuum energy of the theory will have
an analytic expression of the form $m_{q}^{4}F(\log(\frac{m_{q}}{\mu})).$
\ \ In particular for the two loop expression being considered, the function
$F$ \ is expected, at first sight to be a second order polynomial in
$\log(\frac{m_{q}}{\mu})$ . \ \ \ However, the enhanced convergence properties
\ introduced by the high dimension of the parameter $S$ allows to reduce the
dependence to a linear one. To see this property, let us consider both, the
gluon and the quark, propagators as decomposed in their free propagator parts
\ plus a contribution dependent on the condensate. After this, let us also
decompose any of the two loop graphs in the superposition shown in Fig.
\ref{twoloop}, in the set of topologically identical graphs, obtained by
expanding the product of all the internal propagators defining them. Let us
note now the \ fact that, normally \ each divergent loop integration\ adds \ a
pole in to the considered two loop integrals. Then, it can be seen that thanks
to the high \ dimension (of value equal to $3$) of the constant $S,$ it
follows that the only term in which a second order pole in $\epsilon$ can
arise, is the one in which the condensate dependent parts of the propagators
are not appearing. That is, in the graphs of the original mass less QCD.
\ Thus, there is no condensate dependent terms showing a second order pole.
But, the squared logarithm $(\log(\frac{m_{q}}{\mu}))^{2}$ can only arise from
such terms. Thus there are not $(\log(\frac{m_{q}}{\mu}))^{2}$ \ dependent
terms in the considered \ two loop diagrams. This rule explains why they were
no logarithmic terms in the one loop approximation \cite{jhep,epjc2} . \ That
is, the appearance of any condensate dependent component of the propagator in
the considered set of expanded graph makes convergent one of the two loop
integrals and the remaining one can at most produce a linear term in
$\log(\frac{m_{q}}{\mu}).$ \ Further, the presence of two condensate dependent
components of the propagators in two independent loops, leads to the
convergence of the graph, which in these cases shows a simple $m_{q}^{4}$
dependence after removing the dimensional regularization \ \ Thus, the leading
logarithmic correction in the considered situation is linear in $\log(\frac
{m}{\mu}).$ \begin{figure}[h]
%\centering
%
\includegraphics[width=8cm]{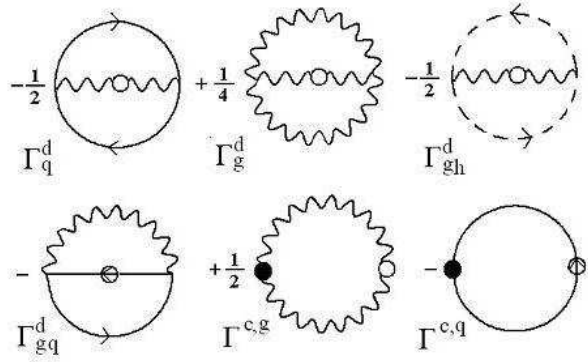}\caption{ The diagram shows the only the
divergent terms which arise after expanding the product of all the
\textit{dressed} quark and gluon propagators in the two diagrams of Fig.
\ref{twoloop}, when they are expressed as the sum of the free propagator plus
the condensate dependent part. }%
\label{lograf}%
\end{figure}

\ The above mentioned properties permit to conclude that the only graphs in
the above described expansion, \ which may contribute to the leading logarithm
correction are the one shown in Fig \ref{lograf}. Let us describe in what
follows the evaluation of the contribution of each of such diagrams to the
logarithmic correction.

The diagram $\Gamma_{q}^{d}$ corresponds to a line of a condensate dependent
gluon propagator connected to the external vertices of the quark loop
contribution $\Pi_{q}^{\mu\nu},$ to the total one loop gluon polarization
operator $\Pi^{\mu\nu}$\ of\ usual mass less QCD. \ The expression for
$\Pi_{q}^{\mu\nu}$can be explicitly evaluated following the conventions in
Ref. \cite{muta}, and $\Gamma_{q}^{d}$ \ takes the form%
\begin{align}
\Gamma_{q}^{d}  &  =\frac{1}{2}\int\frac{dq^{D}}{(2\pi)^{D}i}D_{\mu\nu
}^{(1)ab}(q)\Pi_{q}^{ba\mu\nu}(q)\\
&  =\frac{1}{2}\text{ }a\mu^{4-D}\text{ }g^{2}(N^{2}-1)(D-2)\int\frac{dq^{D}%
}{(2\pi)^{D}i}\frac{J_{1}(-q^{2},\epsilon)}{(q^{2})^{3}+a}\nonumber\\
J_{1}(-q^{2},\epsilon)  &  =\frac{1}{(4\pi)^{\frac{D}{2}}}(-q^{2})^{\frac
{D}{2}-2}\times\frac{\Gamma(2-\frac{D}{2})\Gamma^{2}(\frac{D}{2}-1)}%
{\Gamma(D-2)},\\
\Pi_{q}^{ba\mu\nu}(q)  &  =-\frac{g^{2}}{2}\delta^{ab}\int\frac{dq^{D}}%
{(2\pi)^{D}i}tr_{s}[\gamma^{\mu}G^{(0)}(p+q)\gamma^{\nu}G^{(0)}(p)].
\label{pimunu0}%
\end{align}

The term $\Gamma_{q}^{d}$ \ is associated to a line of \ a condensate
dependent gluon propagator attached to the ends of the gluon loop $\Pi
_{g}^{\mu\nu}$contribution to the one loop polarization operator $\Pi^{\mu\nu
}$ . This quantity also can be directly evaluated to write for this term \
\begin{align}
\Gamma_{g}^{d}  &  =\frac{1}{2}\int\frac{dq^{D}}{(2\pi)^{D}i}(D_{\mu\nu
}^{^{(1)}ab}(q)+D_{\mu\nu}^{(2)ab}(q))\Pi_{g}^{ba\mu\nu}(q)\nonumber\\
&  =-\frac{a\text{ }\mu^{4-D}g^{2}}{8(D-1)}N(N^{2}-1)\int\frac{dq^{D}}%
{(2\pi)^{D}i}\frac{J_{1}(-q^{2},\epsilon)g^{\mu\nu}}{q2((q^{2})^{3}+a)}%
\times\nonumber\\
&  ((6D-5)q^{2}g_{\mu\nu}+(6D-7)q_{\mu}q_{\nu})+\nonumber\\
&  \frac{1}{2}\int\frac{dq^{D}}{(2\pi)^{D}i}D_{\mu\nu}^{(2)ab}(q)\Pi
_{g}^{ba\mu\nu}(q),\label{gammadg}\\
\Pi_{g}^{ba\mu\nu}(q)  &  =\frac{1}{2}\int\frac{dk^{D}}{(2\pi)^{D}%
i}(-ig)f^{acd}V_{\mu\lambda\rho}(-q,k+q,-k)\frac{g^{\lambda\kappa}}{(q+k)^{2}%
}\times\label{pimunug0}\\
&  \frac{g^{\rho\sigma}}{k^{2}}(-i)f^{bdc}V_{\nu\sigma\kappa}%
(q,k,-k-q),\nonumber
\end{align}
where the function $V_{\nu\sigma\kappa}$ defines the three legs gluon vertex
as in Ref. \cite{muta}. \

Similarly, the term $\Gamma_{gh}^{d}$, is associated to the ghost loop
contribution to $\Pi^{\mu\nu}$, can be evaluated \ in the form \
\begin{align}
\Gamma_{gh}^{d}  &  =\frac{1}{2}\int\frac{dq^{D}}{(2\pi)^{D}i}(D_{\mu\nu
}^{(1)ab}(q)+D_{\mu\nu}^{(2)ab}(q))\Pi_{gh}^{ba\mu\nu}(q)\nonumber\\
&  =\frac{a\text{ }\mu^{4-D}g^{2}}{8(D-1)}N(N^{2}-1)\int\frac{dq^{D}}%
{(2\pi)^{D}i}\frac{J_{1}(-q^{2},\epsilon)\text{ \ }g_{\mu\nu}}{q^{2}%
((q^{2})^{3}+a)}\times\nonumber\\
&  ((2-D)q^{\mu}q^{\nu}-q^{2}g^{\mu\nu})+\frac{1}{2}\int\frac{dq^{D}}%
{(2\pi)^{D}i}D_{\mu\nu}^{(2)ab}(q)\Pi_{gh}^{ba\mu\nu}(q),\label{gammagh}\\
\Pi_{gh}^{aa^{\prime}\mu\nu}(q)  &  =-\int\frac{dq^{D}}{(2\pi)^{D}i}%
(-g^{2})\frac{p(p+q)}{p^{2}(p+q)^{2}}f^{abc}f^{a^{\prime}cb}. \label{pimunugh}%
\end{align}

Note that the part associated to the pure longitudinal condensate dependent
propagator$\ D_{\mu\nu}^{(2)ab}$ in Eqs. (\ref{gammadg}) and (\ref{gammagh})
cancels after adding the terms $\Gamma_{gh}^{d}$ and $\Gamma_{g}^{d},$ thanks
to the transversality condition satisfied by the sum of the gluon and ghost
loops terms of the one loop self-energy $\Pi^{\mu\nu}$. This property is
directly seen after adding $\Gamma_{gh}^{d}$ and $\Gamma_{g}^{d}$ which result
can be written as
\begin{align*}
\Gamma_{g}^{d}+\Gamma_{gh}^{d}  &  =\frac{1}{2}\int\frac{dq^{D}}{(2\pi)^{D}%
i}D_{\mu\nu}^{(1)ab}(q)(\Pi_{g}^{ba\mu\nu}(q)+\Pi_{gh}^{ba\mu\nu}(q))\\
&  =\frac{a\text{ }g^{2}}{8(D-1)}N(N^{2}-1)(6D-4)\int\frac{dq^{D}}{(2\pi
)^{D}i}\frac{J_{1}(-q^{2},\epsilon)g^{\mu\nu}}{q^{2}((q^{2})^{3}+a)}\times\\
&  (q^{2}g_{\mu\nu}-q_{\mu}q_{\nu})
\end{align*}
Then, the \ transversal tensor appearing inside the integral eliminates the
purely longitudinal component $D_{\mu\nu}^{(2)ab}$ when added to $D_{\mu\nu
}^{(1)ab}$.

Further, the last of the diagrams not being associated to counterterms,
$\Gamma_{gq}^{d},$\ is related with a line of quark condensate dependent
propagator connected to the external vertices of the \ one loop contribution
to the quark self-energy in usual mass less QCD. After evaluating the one loop
quark self-energy $\Sigma_{sr}^{ij}$ , this term can be written as follows
\begin{align}
\Gamma_{gq}^{d}  &  =-\int\frac{dq^{D}}{(2\pi)^{D}i}G_{rs}^{ji}(q)\Sigma
_{sr}^{ij}(q)\nonumber\\
&  =-\text{ }\mu^{4-D}g^{2}(2-D)(N^{2}-1)\int\frac{dq^{D}}{(2\pi)^{D}i}%
\frac{(q^{2})^{3}J_{1}(-q^{2},\epsilon)}{((q^{2})^{3}+a)},\label{gammagq}\\
\Sigma_{sr}^{ij}(q)  &  =\int\frac{dp^{D}}{(2\pi)^{D}i}\gamma_{\mu}%
^{r^{\prime}r}T_{j^{\prime}j}^{a}G^{(0)s^{\prime}r^{\prime}}(p)\frac
{1}{(q-p)^{2}}g\text{ }\gamma^{\mu,ss^{\prime}}T_{ij^{\prime}}^{a}.
\label{sigma0}%
\end{align}

Next, let us write the expressions for the \ diagrams related with the
counterterms. \ The values for the renormalization constants will be chosen as
coinciding with the ones in QCD as evaluated in Ref. \ \cite{muta}. \ The
gluon counterterm loop will be decomposed in the two components being
associated to the substractions $\Gamma_{g}^{c,g}$ and $\Gamma_{q}^{c,g}$
which makes finite the gluon and quark loops contribution to the polarization
operator respectively, in standard QCD. These quantities can be written \ in
the form \
\begin{align}
\Gamma^{c,g}  &  =\Gamma_{g}^{c,g}+\Gamma_{q}^{c,g},\label{gammacg}\\
\Gamma_{g}^{c,g}  &  =\frac{a\text{ }\mu^{4-D}}{2}(D-1)(N^{2}-1)\delta
Z_{3}^{g}\int\frac{dq^{D}}{(2\pi)^{D}i}\frac{1}{((q^{2})^{3}+a)}%
,\label{gammacgg}\\
\Gamma_{q}^{c,g}  &  =\frac{a\text{ }\mu^{4-D}}{2}(D-1)(N^{2}-1)\delta
Z_{3}^{q}\int\frac{dq^{D}}{(2\pi)^{D}i}\frac{1}{((q^{2})^{3}+a)}%
,\label{gammacgq}\\
Z_{3}  &  =1+\delta^{g}Z_{3}+\delta Z_{3}^{f},\label{Z3}\\
\delta Z_{3}^{g}  &  =(\frac{g_{0}}{4\pi})^{2}\frac{1}{\epsilon}\frac{1}%
{2}C_{G}(\frac{10}{3}),\label{dZ3g}\\
\delta Z_{3}^{q}  &  =-(\frac{g_{0}}{4\pi})^{2}\frac{1}{\epsilon}\frac{4}%
{3}T_{R},\label{dZ3q}\\
T_{R}  &  =\frac{1}{2},\text{ \ \ }C_{G}=N.\text{\ } \label{TRCG}%
\end{align}

In the case of the quark loop counterterm contribution in Fig. \ref{lograf} ,
the \ \ following expression can be obtained%
\begin{align}
\Gamma^{c,q}  &  =4N(Z_{2}-1)\int\frac{dq^{D}}{(2\pi)^{D}i}\frac{(q^{2})^{3}%
}{((q^{2})^{3}+a)},\label{gammacq}\\
Z_{2}  &  =1-(\frac{g}{4\pi})^{2}\frac{C_{F}}{\epsilon},\label{Z2}\\
\text{\ \ }C_{F}  &  =\frac{N^{2}-1}{2N}. \label{CF}%
\end{align}

Note that the mass renormalization parameter has been taken equal to zero in a
first instance to check whether the same renormalization constants of the
massless QCD\ are able to eliminate the infinities in the \ considered
approximation. However, since we are \ investigating the mass generation an
improved selection could will be performed after better understanding the
problem. These possibilities will be considered \ in the extension of the work.

\subsection{Leading logarithm corrections}

The evaluations to be considered in this section will be done in Euclidean
space and by selecting only the real part of the integrals. \ That is, we will
perform the substitution \ \ $p_{0}\rightarrow i$ $p_{4}$ (with $p_{4}$
\ real) in the integrals an take the real part of it. \ \ This means that we
will be effectively calculating the real part of the zero temperature
Thermodynamical Potential of the system. A difference between this quantity
and the Effective Action in Minkowski \ space could appear when it is
attempted to perform\ a continuous deformation of the integration path, in
order to implement the above mentioned simple substitution. \ We will not
analyze this effect in this paper, since it is related with possible
instabilities of the systems showing such non vanishing differences. Those
instabilities \ can be expected to arise in this "one condensate" calculation,
by example, if not only \ one, but few quark condensates are needed to be
dynamically generated in order to arrive at the real ground state of the system.

After adding $\Gamma_{q}^{d}$ $\ $and $\Gamma_{q}^{c,g},$ the following
expression, which explicitly shows the\ dependence on the dimension $D,$ can
be written%
\begin{align*}
\text{ \ }\Gamma_{q}^{d}+\Gamma_{q}^{c,g}  &  =-(N^{2}-1)\int_{0}^{\infty
}dq_{0}\int_{0}^{\infty}dr\frac{1}{(q^{2}+i\delta)^{3}+1}\frac{r^{D-2}}%
{(2\pi)^{D}i}\frac{2\pi^{D-1}}{\Gamma(\frac{D-1}{2})}\times\\
&  {\Huge (}-\frac{(D-2)\Gamma(2-\frac{D}{2})\Gamma^{2}(\frac{D}{2}-1)}%
{2(4\pi)^{\frac{D}{2}}\Gamma(D-2)}(-q^{2})^{\frac{D}{2}-2}a^{\frac{D}{3}%
-\frac{2}{3}}\mu^{4\epsilon}g_{0}^{2}+\\
&  +\frac{(D-1)}{2}(\frac{g_{0}}{4\pi})^{2}\mu^{2\epsilon}a^{\frac{D}{6}}%
\frac{4}{3}T_{R}\frac{1}{\epsilon}{\Huge )},\text{ \ \ \ }\\
q^{2}  &  =q_{0}^{2}-r^{2},\text{ }\epsilon=2-\frac{D}{2}.
\end{align*}

Taking the limit $\epsilon\rightarrow0$, the above formula allows to verify
that the \ usual counterterm of \ mass less QCD \ cancels the divergence of
this Effective Action contribution. After the Wick substitution $q_{0}%
\rightarrow i$ $q_{4}$\ (without considering the residues at the poles as
described above) \ the real part of the effective potential, \ is given by the
above expression \ in which the integral is taken in the principal value
sense. The explicit evaluation \ gives for the contribution of this term to
the leading logarithm correction, the result%
\begin{align}
V_{a}(m_{q})  &  =-\lim_{\epsilon\rightarrow0}\operatorname{Re}[\Gamma_{q}%
^{d}+\Gamma_{q}^{c,g}]\nonumber\\
&  =\frac{\pi}{2}(N^{2}-1)\frac{g_{0}^{2}\text{ }m_{q}^{4}}{384\text{
}6^{\frac{1}{3}}\pi^{5}}PP\int_{0}^{\infty}dx\text{ }x^{3}\frac{1}{-x^{6}%
+1}\times\nonumber\\
&  (3\log(x^{2})+6\log(\frac{m_{q}}{\mu})-3\gamma-5+\log(\frac{3}{256\text{
}\pi^{3}}))\nonumber\\
&  =\frac{g_{0}^{2}m_{q}^{4}}{1728\sqrt[3]{6}\pi^{3}}\left(  \sqrt{3}\left(
6\log\left(  \frac{m_{q}}{\mu}\right)  +\log\left(  \frac{3}{256\pi^{3}%
}\right)  +3\gamma-5\right)  +4\pi\right)  . \label{v(m)a}%
\end{align}

Adding the terms $\Gamma_{g}^{d}$ and $\Gamma_{gh}^{d}$ with $\ $their
corresponding counterterm contribution $\Gamma_{g}^{c,g}$ the resulting
dependence on the dimension $D$ \ is given by the formula%
\begin{align*}
\Gamma_{g}^{d}+\Gamma_{gh}^{d}+\Gamma_{g}^{c,g}  &  =-2(N^{2}-1)\int
_{0}^{\infty}dq_{0}\int_{0}^{\infty}dr\text{ }\frac{r^{D-2}}{(2\pi)^{D}i}%
\frac{1}{(q^{2}+i\delta)^{3}+1}\frac{2\pi^{D-1}}{\Gamma(\frac{D-1}{2})}%
\times\\
&  {\Huge (}\frac{\Gamma(2-\frac{D}{2})\Gamma^{2}(\frac{D}{2}-1)}{(4\pi
^{\frac{D}{2}})\Gamma(D-2)}\frac{N\text{ }(3D-2)}{4(D-1)}(-q^{2})^{\frac{D}%
{2}-2}a^{\frac{D}{3}-\frac{2}{3}}\mu^{4\epsilon}g_{0}^{2}-\\
&  \frac{1}{4}(\frac{g_{0}}{4\pi})^{2}\mu^{2\epsilon}a^{\frac{D}{6}}\frac
{1}{\epsilon}C_{G}\frac{10}{3}{\Huge )},\text{ \ \ \ \ }q^{2}=q_{0}^{2}-r^{2},
\end{align*}
in which again, the $\epsilon\rightarrow0$ \ limit shows that divergences are
canceled by the original one loop counterterms of mass less QCD. As before,
after the Wick substitution, for the real part of this contribution to the
potential, the result \ can be expressed as follows
\begin{align}
V_{b}(m_{q})  &  =-\lim_{\epsilon->0}\operatorname{Re}[\Gamma_{g}^{d}%
+\Gamma_{gh}^{d}+\Gamma_{g}^{c,g}]\nonumber\\
&  -\frac{\pi}{2}(N^{2}-1)\frac{g_{0}^{2}\text{ }m_{q}^{4}}{256\text{
}6^{\frac{1}{3}}\pi^{5}}PP\int_{0}^{\infty}dx\text{ }x^{3}\frac{1}{-x^{6}%
+1}\times\nonumber\\
&  (15\log(x^{2})+30\log(\frac{m_{q}}{\mu})+15\gamma-31-40\log2+\log
(\frac{243}{\pi^{15}}))\nonumber\\
&  =-\frac{g_{0}^{2}m_{q}^{4}}{1152\sqrt[3]{6}\pi^{3}}{\Large (}\sqrt
{3}{\Large (}30\log\left(  \frac{m_{q}}{\mu}\right)  -15\log(\pi
)+\log(243)\nonumber\\
&  \text{ \ \ \ \ }-40\log(2)+15\gamma-31{\Large )}+20\pi{\Large ).}
\label{v(m)b}%
\end{align}

Finally, the\ quantities $\Gamma_{gq}^{d}$ $\ $ and $\Gamma^{c,q}$ , after to
be added can be expressed in the following form as explicit functions of the
dimension $D$

\begin{figure}[h]
%\centering
%
\includegraphics[width=8cm]{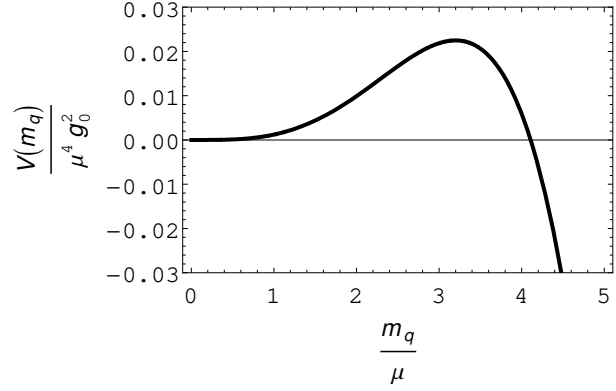}\caption{ The figure illustrates the
dependence of the logarithmic in the condensate contribution to the effective
potential. The curve indicates an instability of the system in the
approximation being considered, under the generation of large condensate
values.Thus, the approximation adopted is yet insufficient to detect the
existence of a stable ground state to which the system would relax. However,
the natural appearance of squared logarithms of the condensate in the next
three loop approximation makes feasible that the stability can arise after
considering three loop corrections. }%
\label{potential}%
\end{figure}%
\begin{align*}
\Gamma_{gq}^{d}+\Gamma^{c,q}  &  =-g^{2}C_{F}N(2-D)J_{1}(1,\epsilon)\frac
{2\pi^{\frac{D}{2}}}{\Gamma(\frac{D}{2})}\frac{S^{\frac{^{2}}{3}D-\frac{4}{3}%
}}{(2\pi)^{D}}\frac{\pi}{3}\sec(\frac{\pi}{6}(1-2D))+\\
&  +\frac{4C_{F}N}{\epsilon}(\frac{g_{0}}{4\pi})^{2}\frac{2\pi^{\frac{D}{2}}%
}{\Gamma(\frac{D}{2})}\frac{(S^{2})^{\frac{D}{6}}}{2(2\pi)^{D}}\frac{\pi}%
{3}\sec(\frac{\pi}{6}(1-2\epsilon)),
\end{align*}
where as before, the standard counterterm is able to \ cancel the divergence.
Now, after performing the Wick substitution \ the contribution to the
potential in the limit \ $\epsilon\rightarrow0,$ takes the expression
\begin{align}
V_{c}(m_{q})  &  =-\lim_{\epsilon->0}\operatorname{Re}(\Gamma_{gf}^{d}%
+\Gamma^{c,q})\nonumber\\
&  =-\frac{g_{0}^{2}m_{q}^{4}}{216\pi^{3}}\left(  \sqrt{3}\left(  6\log
(m_{q})-6\log(\mu)+\log\left(  \frac{1}{64\pi^{3}}\right)  +3\gamma-3\right)
+\pi\right)  . \label{v(m)c}%
\end{align}

\ Finally, the \ complete leading logarithm correction to the potential has
the form%
\begin{align}
V(m_{q})  &  =V_{a}(m_{q})+V_{b}(m_{q})+V_{c}(m_{q})\nonumber\\
&  \cong g_{0}^{2}\text{ }m_{q}^{4}{\large (}-C_{1}\text{ }\log(\frac{m_{q}%
}{\mu})+C_{2}{\large ),}\label{v(m)}\\
C_{1}  &  =8.5788\times10^{-4},\text{ \ }C_{2}=1.21224\times10^{-3}.\nonumber
\end{align}

\ This component of the potential is plotted as a function of the $quark$
$mass$ in Fig. \ref{potential}. \ The picture illustrates the instability
under the generation of \ large quark condensate. However, the considered
approximation, does not \ furnishes yet \ a \ minimum around which the system
can stabilize.

However, such a minimum can be a natural consequence in a three loop level at
which ($\log(m_{q}))^{2}$ \ corrections should appear, whenever the net sign
turns to be the appropriate one. \ This \ question is expected to be
considered elsewhere.

\section{The Higgs particle as a $t\overline{t}$ meson}

\begin{figure}[h]
%\centering
%
\includegraphics[width=8cm]{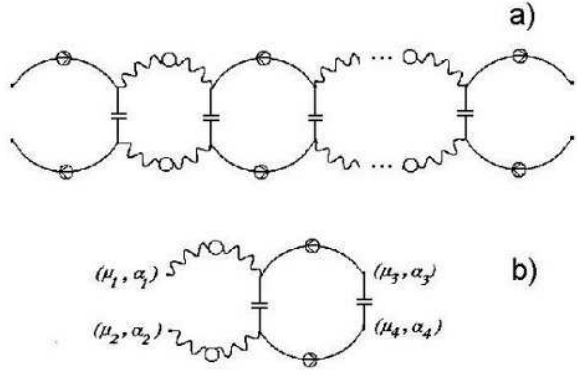}\caption{ Figure a) shows the diagram
associated to general power correction in the geometric series defining the
ladder approximation to the two particle Green function. The ladder
approximation refers to the zeroth order in the coupling constant
approximation. That is, it only includes the condensate dependent vertex.
Figure b) illustrates the repeated diagram in terms of which the geometric
series was evaluated and the indices convention. }%
\label{G2}%
\end{figure}

In this section we will assume that the next three loop corrections will be
able to produce a minimum in the potential and that this minimum can be fixed
to occur at the top quark mass value $\ m_{q}=173$ $GeV$ . \ Let us \ discuss
below an evaluation of the singularities of the $t$ and $\overline{t}$ two
particle propagator $G_{\mu_{1},\mu_{2};\mu_{3},\mu_{4}}^{a_{1},a_{2}%
;a_{3},a_{4}}(p_{1},p_{2})$ after to be contracted in its color and
spinor indices at\ the input and output pairs of legs. \ This
contracted propagator corresponds to zero color and spin channel of
the $t$ and $\overline{t}$ quarks. . The calculation will be
considered in the ladder approximation \ in the condensate dependent
vertex \ and the zero order in the coupling constant. The
Fig.\ref{G2} a) shows one general term of the geometric series
defining the ladder approximation. Fig.\ref{G2}b) illustrates the
basic diagram which \ repetitive insertion permits generate the
general diagram shown in Fig.\ref{G2} a)\ 

The contribution corresponds to the zeroth order in the coupling
constant $g$ and a ladder approximation of the propagator in the
condensate vertex. \ \ In order to simplify the evaluation we will
consider that the parameter $b$ defining the contribution
$D^{(2)}(p)$ of the gluon propagator is much smaller than the
parameter $a$ associated to $D^{(1)}(p)$ in the proportion $\
b/a=9/128$ \ \ Thus, the terms $D^{(2)}(p)$ will be disregarded
which simplifies the discussion due to the simpler Lorentz structure
of \ $D^{(1)}(p)$.

The \ evaluation of the repetitive block defining the ladder approximation
(illustrated in Fig.\ref{G2}b) \ leads to the \ expression \
\begin{align*}
T_{\mu_{1},\mu_{2};\mu_{3},\mu_{4}}^{a_{1},a_{2};a_{3},a_{4}}(p_{1},p_{2})  &
=-\frac{2(p_{1}^{2}p_{2}^{2})^{2}}{((p_{1}^{2})^{3}+a)((p_{2}^{2})^{3}%
+a)}(\frac{g^{2}C_{q}}{(2\pi)^{D}})^{2}Tr_{c}{\large (}T^{a_{1}}T^{a_{2}%
}T^{a_{4}}T^{a_{3}}{\large )}\times\\
&  Tr_{s}{\large (}\gamma_{\mu_{1}}\gamma_{\mu_{2}}G(-p_{2})\gamma^{\mu_{4}%
}\gamma^{\mu_{3}}G(p_{1}){\large ).}%
\end{align*}
However, after contracting this expression under the input color indices
$a_{1}$ and $a_{2}$ and the spinor ones $\mu_{1}$ and $\mu_{2}$ \ and
employing the relations
\begin{align*}
Tr_{c}{\large (}T^{a_{1}}T^{a_{1}}T^{a_{4}}T^{a_{3}}{\large )}  &
=\frac{C_{F}}{2}\delta^{a_{3}a_{4}},\\
Tr_{s}{\large (}\gamma_{\mu_{1}}\gamma^{\mu_{1}}G(-p_{2})\gamma^{\mu_{4}%
}\gamma^{\mu_{3}}G(p_{1}){\large )}  &  =\frac{4D(-p_{1}.p_{2}+\frac{S^{2}%
}{p_{1}^{2}p_{2}^{2}})}{(p_{1}^{2}-\frac{S^{2}}{(p_{1}^{2})^{2}})(p_{2}%
^{2}-\frac{S^{2}}{(p_{2}^{2})^{2}})}g^{\mu_{3}\mu_{4}},
\end{align*}
it follows that the dependence on the output color and spinor indices get the
simple structure
\begin{align*}
g^{\mu_{1}\mu_{2}}T_{\mu_{1},\mu_{2};\mu_{3},\mu_{4}}^{a_{1},a_{1};a_{3}%
,a_{4}}(p_{1},p_{2})  &  =-\frac{4DC_{F}(p_{1}^{2}p_{2}^{2})^{2}}{((p_{1}%
^{2})^{3}+a)((p_{2}^{2})^{3}+a)}(\frac{g^{2}C_{q}}{(2\pi)^{D}})^{2}\times\\
&  \frac{(-p_{1}.p_{2}+\frac{S^{2}}{p_{1}^{2}p_{2}^{2}})}{(p_{1}^{2}%
-\frac{S^{2}}{(p_{1}^{2})^{2}})(p_{2}^{2}-\frac{S^{2}}{(p_{2}^{2})^{2}}%
)}\delta^{a_{3}a_{4}}g^{\mu_{3}\mu_{4}},\\
&  =T(p_{1},p_{2})\delta^{a_{3}a_{4}}g^{\mu_{3}\mu_{4}}.
\end{align*}
Therefore, all the set of intermediate color and spinor indices in the
successive blocks in the ladder expansion also contracts. \ This property
allows to sum the geometric series associated with the ladder approximation to
write the considered Green function in the following form
\[
g^{\mu_{1}\mu_{2}}G_{\mu_{1},\mu_{2};\mu_{3},\mu_{4}}^{a_{1},a_{1};a_{3}%
,a_{3}}(p_{1},p_{2})g^{\mu_{3}\mu4}=\frac{1}{1-T(p_{1},p_{2})}F(p_{1},p_{2})
\]
where $F$ is given by the same function $T$ defined above, after being
multiplied by a function of the momenta defined by the external lines of $G$.
Henceforth, the singularities of this propagator in the considered
approximation are defined by the zeroes of the denominator
\begin{align*}
0  &  =1-T(p_{1},p_{2})\\
&  =1+(\frac{g^{2}C_{q}}{(2\pi)^{D}})^{2}\frac{4DC_{F}(p_{1}^{2}p_{2}^{2}%
)^{2}}{((p_{1}^{2})^{3}+a)((p_{2}^{2})^{3}+a)}\times\\
&  \frac{(-p_{1}.p_{2}+\frac{S^{2}}{p_{1}^{2}p_{2}^{2}})}{(p_{1}^{2}%
-\frac{S^{2}}{(p_{1}^{2})^{2}})(p_{2}^{2}-\frac{S^{2}}{(p_{2}^{2})^{2}})}.
\end{align*}
This relation can be now expressed in terms of the center of mass momentum $p$
and the relative momentum $q,$ defined by
\begin{align*}
p  &  =p_{1}+p_{2}\\
q  &  =p_{1}-p_{2},
\end{align*}
in the following form
\begin{align}
1  &  =-\frac{S^{2}}{C_{F}}\frac{((p+q^{2})^{2}-4(p.q)^{2})^{4}}{4^{4}%
(\frac{(p^{2}+2p.q+q^{2})^{3}}{2^{6}}+\frac{4}{3}S^{2})(\frac{(p^{2}%
-2p.q+q^{2})^{3}}{2^{6}}+\frac{4}{3}S^{2})}\times\nonumber\\
&  \frac{(-\frac{1}{4}(p^{2}-q^{2})+\frac{16S^{2}}{(p^{2}+q^{2})^{2}%
-4(p.q)^{2}})}{(\frac{(p^{2}+2p.q+q^{2})^{3}}{2^{6}}-S^{2})\frac
{(p^{2}-2p.q+q^{2})^{3}}{2^{6}}-S^{2}}. \label{dispersion}%
\end{align}

Now, without loss of generality, by selecting the reference frame
\ appropriately, these vectors can be expressed in terms of three parameters
\ as
\begin{align*}
p  &  =(m,0,0,0),\\
q  &  =(q_{0},\mathbf{q,}0,0).
\end{align*}

\begin{figure}[h]
%\centering
%
\includegraphics[width=8cm]{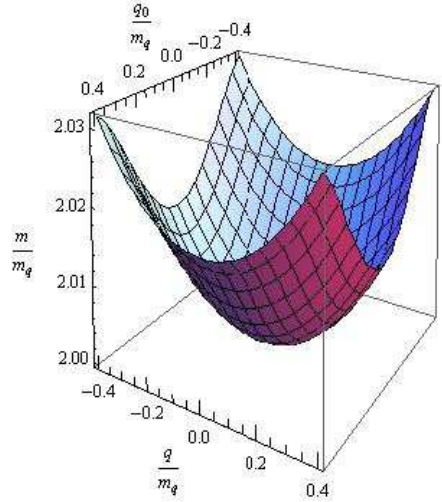}\caption{The figure illustrates the
particle anti-particle continuous spectrum of excitations associated to the
evaluated two particle Green function correction. Note that the further
inclusion of the Bethe-Salpeter gluon kernel, can produce bound state below
the mass threshold of the $t$ and $\overline{t}$ generation in the $top$ quark
detection experiments. }%
\label{band1}%
\end{figure}

One branch of solutions of the dispersion relation (\ref{dispersion}),
expressed as the values of the $center$ $of$ $mass$ $energy$ \ $m$ as a
function of the component of the relative momentum $q_{0}$ and $\mathbf{q,}$
is illustrated in Fig. \ \ref{band1}. \ The plot shows a continuous spectrum
\ with a mass threshold equal to twice the $top$ quark mass $m_{q}$.\ This
results is a natural one in the here considered approximation, in which the
order $g^{2}$ gluon interacting kernel has not been considered. \ The obtained
spectrum \ also suggests that the Higgs particle in the here considered
scheme, should correspond to a possibly existing short living $t\overline{t}$
bound state \ appearing below the $2m_{p}=346$ $\ GeV$ \ mass threshold. This
meson can be expected to be described by the present picture after \ also
introducing .the gluon kernel into the Bethe-Salpeter equation associated with
the here examined two \ particle Green function. \ \begin{figure}[h]
%\centering
%
\includegraphics[width=8cm]{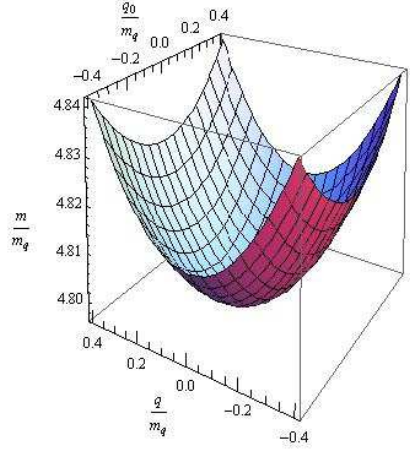}\caption{ A second continuous spectrum
branch of $t\overline{t}$ excitations following from the considered zeroth
order in the coupling two points Green function. Note that it shows a
threshold near $4.8m_{q}.$ }%
\label{band2}%
\end{figure}The dispersion relation (\ref{dispersion}) shows another threshold
for continuous excitation. It is shown in Fig. \ref{band2}. The mass gap in
this case is at a value $4.8$ $m_{q}$= $830.4$ $GeV$ .\ Thus, after including
the color binding kernel an excited state of the $t\overline{t}$\ meson could
exists near this higher threshold.

\section{Summary}

In this work a functional integral representation has been presented
which \ promises to be helpful in the study of ground states showing
a quark condensate. The analysis suggests that it could be also
technically useful in applications to systems \ showing
superconductivity. Although the discussion is only restricted to the
presence of a single quark condensate, the extension to the case of
the inclusion of other quark and gluon condensates seems to be
feasible. \ \ The new vertex opens possibilities for start the
construction of an alternative to the SM in which the mass and CKM
matrices are determined \ by \ the condensation of the various
quarks and leptons. Thus, the results suggest that the instability
of the massless QCD could be the driving force generating the
hierarchical particle mass spectrum, though a kind of generalized
Nambu-Jona Lasinio mechanism.    The procedure is employed here to \
study the possibility for the existence of a stable ground state
showing a quark condensate, through a two loop evaluation \ of \ the
vacuum energy. For this purpose, the leading logarithm dependence \
on the quark condensate parameter of the two loop correction to the
Effective Potential is calculated. The result indicates that the
system is unstable under the generation of large values of the
condensate,  but a  stabilization point in the dependence is not
following in the considered approximation. However, the expected \
to appear quadratic terms in $\log(\frac{m_{q}}{\mu})$ \ at the
three loop level, \ \ have the chance of determining \ a minimum of
the potential. \ \ In addition, improved evaluations of the vacuum
energy, not being based in the recourse of \ simply substituting the
mass less propagators by approximate $dressed$ ones in the usual
loop expansion,\ can also be of help. These possibilities are
expected to be considered in the extension of the work.

The expansion is also employed to evaluate the two particle \ Green
function associated to a color and scalar singlet channel in the
ladder approximation in terms of the condensate dependent vertex.
Assumed that the quark mass can be fixed to the observed value, the
evaluation in this simple approximation shows a continuous spectrum
of excitation \ above a mass threshold equal to two times the $top$
quark mass. \ Therefore, following the idea of the old \ $top$
condensate models, in which the role of the Higgs is played by the
$t\overline{t}$ condensate, the discussion indicates that the Higgs
could correspond to a $t\overline{t}$ bound state meson which could
be detected below the threshold \ found in the experiments for
finding  the top quark. \ The study of such bound states after
incorporating the \ gluon kernel in the Bethe-Salpeter equation is
expected to be considered in extending the work.

Before ending,  let us comment on two important issues both
connected with the extension of the work. The first of them  is
related with the possibility that the mechanism under discussion can
\ describe the $top$ quark and $\Lambda_{QCD}$ mass scales at the
same time. \ This opportunity is suggested after noticing two main
properties: a) The logarithmic dependence of the effective potential
on the quark mass $m_{q}$ \ and b) \ The presence of terms in the
effective potential of the form \ \ $-|a|$ $m_{q}^{4},$ reflecting
the instability.  \ To see it, let us consider the potential in two
loops (at more loops, higher powers of the
logarithm will appear) as written in the form\ %
\begin{align*}
V(m_{q}) &  =-|a|\text{ }m_{q}^{4}\text{ }+b\text{ }m_{q}^{4}\text{ }\log
\frac{m_{q}}{\mu}\\
&  =|b|\text{ }m_{q}^{4}\text{ }\log(\frac{m_{q}}{\exp(\frac{|a|}{b})\mu}),
\end{align*}
where it is assumed that after the exact evaluation of the finite
terms (having a dependence of the form $m_{q}^{4})$, the resulting
coefficient shows  the negative value $-|a|$ \ \ Further assume that
the extremum of the potential as a function of $m_{q}$ , is situated
at the $top$ quark mass $m_{q}=173$ GeV $\ $\ and also that the
dimensional regularization parameter $\mu$ is approximately given by
$\Lambda_{QCD}$ $\approx0.1$ GeV. \ \ Therefore, after finding the
extremum of the potential over $m_{q}$, the
ratio of the coefficients $\frac{|a|}{|b|}$ is estimated as follows %
\begin{align*}
0  & =m_{q}^{3}\text{ }(4\log(\frac{m_{q}}{\exp(\frac{|a|}{b})\mu
})+1)\rightarrow\exp(\frac{|a|}{b})=\exp(\frac{1}{4})\frac{m_{q}}{\mu}\\
\frac{|a|}{|b|}  & =|\log(\exp(\frac{1}{4})\frac{m_{q}}{\mu})|\approx\log
(\exp(\frac{1}{4})\frac{173\text{ GeV}}{0.1\text{ GeV}})=7.70.
\end{align*}
Henceforth, since the instability created by the $QCD$ forces can be
strong, it seems feasible that a not so large ratio value of
$\frac{|a|}{|b|}=7.7$ could arise after evaluating the finite terms,
then giving space for these two widely different scales be both
predicted by the scheme.
\begin{figure}[h]
\includegraphics[width=8cm]{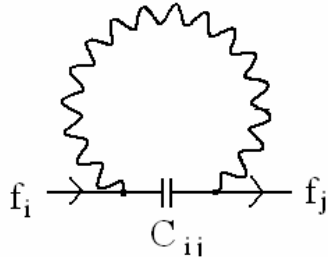}
\caption{The figure show the effective action two legs diagrams
through
 which the generalized theory including all sort of condensates  from the start
  could be able to generate the Yukawa mass and CKM matrices.}%
\label{genmod1}%
\end{figure}
\begin{figure}[h]
\includegraphics[width=8cm]{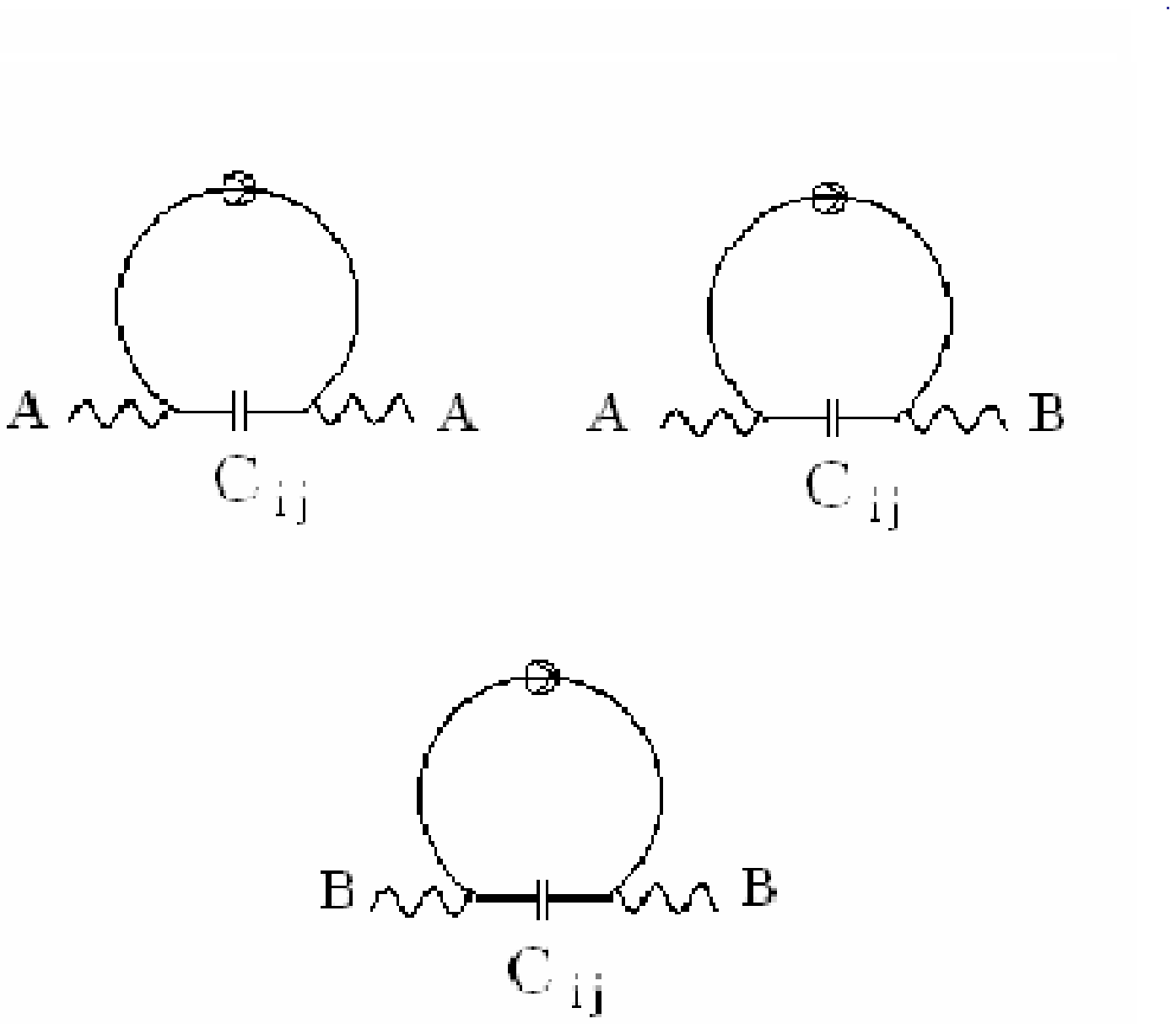}
\caption{The figure illustrates  few diagrams in the effective
action of the generalized theory with two legs of weak interaction
bosons,  which could generate the masses of the $W$ and $Z$ particles.}%
\label{genmod2}%
\end{figure}

 The second point about we which to remark is connected with the
 possibility for constructing a variant of the SM model starting from the
 here presented discussion. This idea may directly come to the mind
 by noticing that the mass terms associated to the quark and leptons in
  a fist approximation for an effective model are associated to the
  graph of the form illustrated in figure \ref{genmod1} showing one condensate
  vertex with two external fermion legs and a massive gluon
  propagator joining the two gluon legs of the vertex. In the
before proposed generalized form of the vertex the indices $f$  can
correspond to quarks or leptons, a fact that lead to idea of fixing
the matrix $C_{f_1,f_2}$ to reproduce the Yukawa mass matrix. Note
that the large mass of the gluon modified propagator should make the
interaction between the input and output fermions short ranged  and
then the vertex effectively produce  a Yukawa term.
 Figure \ref{genmod2} simply intends to show some diagrams of the
 considered generalized model which
could generate the masses for the $W$ and $Z$ particles  starting
from the massless $SU(2)$ and $U(1)$  vector gauge fields of the SM
model.The possibilities indicated above are in some measure
supported by the fact that in the context of the usual top
condensate models, it has been argued that the top anti-top
condensate technically implements the role of the Higgs field.

\begin{acknowledgments}
The authors wish to acknowledge the helpful support received from various
institutions: the Caribbean Network on Quantum Mechanics, Particles and Fields
(Net-35) of the ICTP Office of External Activities (OEA), the "Proyecto
Nacional de Ciencias B\'{a}sicas" (PNCB) of CITMA, Cuba and the Perimeter
Institute for Theoretical Physics, Waterloo, Canada.
\end{acknowledgments}

\end{document}